\documentclass{article}
\usepackage{spconf,amsmath,amssymb,graphicx,booktabs,multirow,array,xcolor,url}
\usepackage{float}
\usepackage{hyperref}

\let\oldthebibliography\thebibliography
\renewcommand{\thebibliography}[1]{%
  \oldthebibliography{#1}%
  \setlength{\itemsep}{0pt}%
  \setlength{\parskip}{0pt}%
  \setlength{\parsep}{0pt}%
}

\title{Beyond the Benchmark: \\ Conditioning, Domains and Failure Modes of
Qwen3-TTS}

\name{\begin{tabular}{@{}c@{}}
      Séverin Baroudi$^{1,2}$ \quad Yanis Labrak$^{3}$ \\
      Petr Motlicek$^{3,4}$
      \quad
      Hervé Bredin$^{5}$
      \quad
      Ricard Marxer$^{1,6}$
      \end{tabular}}
\address{
  $^{1}$Univ Toulon, Aix Marseille Univ, LIS, CNRS, France;  $^{2}$Mila -- Quebec AI Institute \\
  $^{3}$Idiap Research Institute, Martigny, Switzerland;
  $^{4}$Brno University of Technology, Brno, Czech Republic \\
  $^{5}$pyannoteAI, Toulouse, France \quad
  $^{6}$CNRS, ILLS, Montréal, Canada \\[2pt]
  }

\begin{document}
\ninept
\maketitle

\begin{abstract}
Qwen3-TTS reports state-of-the-art zero-shot voice cloning, but its
technical report leaves open how the model should be conditioned, how it
behaves outside read speech, and how it fails. We characterise its two
cloning interfaces, in-context learning (ICL) and X-Vector, against
Chatterbox, XTTS-v2 and F5-TTS on read, meeting, emotional and
air-traffic-control (ATC) speech, using eighteen objective measures, an
LLM judge, a spoofing countermeasure and a listening test. Neither
interface dominates: ICL clones timbre more closely from correctly
transcribed references, X-Vector clones timing, and ICL degrades sharply
when a short reference transcript is wrong. Domain explains four times
more variance than system choice. Qwen3-TTS leads on intelligibility and
pitch register but trails on prosody, and the objective measures rank
systems differently from listeners. An off-the-shelf countermeasure flags its clones no more often than
genuine recordings, and on ATC radio it fails for every system. Code and
audio are released.
\end{abstract}


\section{Introduction}
\label{sec:intro}

Since neural codec language models made speaker-conditioned synthesis
practical \cite{wang2023valle}, zero-shot voice cloning has become a
crowded field of open systems such as XTTS-v2 \cite{casanova2024xtts},
F5-TTS \cite{chen2024f5}, Chatterbox\footnote{\url{github.com/resemble-ai/chatterbox}}  and, most
recently, Qwen3-TTS \cite{qwen3tts}, several reporting human parity on
read-speech benchmarks \cite{chen2024valle2,ju2024naturalspeech3}.
Qwen3-TTS reports state-of-the-art zero-shot cloning and, unusually,
exposes two distinct cloning interfaces. Its technical report, like most
system papers, establishes that the model is strong, but it does not tell a
practitioner which interface to use, how the model behaves away from
clean read speech, or how it fails.

Existing evaluations report ASR word
error rate \cite{radford2023whisper}, speaker-embedding similarity
\cite{snyder2018xvector,wang2023wespeaker} and predicted MOS
\cite{reddy2022dnsmos}, mostly on clean read speech.
These saturate as systems approach human quality, MOS tests are
sensitive to design \cite{lemaguer2024mos}, and none
says which acoustic dimension betrays a clone or how quality changes
across domains. TTSDS2 \cite{minixhofer2025ttsds2} broadens the
objective battery, the human fooling rate
\cite{srinivasavaradhan2025hfr} measures deception directly, audio LLMs
are emerging as judges \cite{chiang2025judges} with only partial human
agreement \cite{zhang2026speechjudge}, and deepfake detection
\cite{yamagishi2021asvspoof,jung2022aasist} asks a different,
machine-centred question. None covers Qwen3-TTS or its two interfaces,
and no single family of evidence suffices alone, so we combine all four
over four domains differing in recording conditions, speaking style and
vocabulary.

Listeners already miss a large share of speech deepfakes \cite{mai2023warning} and often, LLM based judge TTS outputs to be real speakers in about $80\%$ of the time
\cite{barrington2025voiceclones,lavan2025hyperreal}, which fuels
voice-phishing fraud that current detectors struggle to stop
\cite{gengap2025} and has led the EU AI Act to mandate disclosure of
synthetic audio \cite{euaiact2024}.

We therefore ask, how convincing the
five systems are to listeners (Section~\ref{sec:rq1}), whether an
off-the-shelf spoofing detector catches what listeners miss
(Section~\ref{sec:rq1b}), whether Qwen3-TTS should be conditioned on
reference audio and transcript or on a speaker embedding alone
(Section~\ref{sec:rq2}), which domains and inputs make it break down
(Section~\ref{sec:rq3}), which acoustic dimensions are strong or lacking
(Section~\ref{sec:rq4}), and whether automatic measures can stand in for
human judgment (Section~\ref{sec:rq5}).
All code and the generated audio are released publicly.\footnote{\url{beyond-the-benchmark.github.io}}

\section{Experimental Setup}
\label{sec:setup}

\subsection{Domains, systems and cloning conditions}

Four English corpora test sets, covering distinct domains: LibriSpeech \cite{panayotov2015librispeech} (LS), clean (read speech); AMI \cite{carletta2005ami} (far-field spontaneous meetings); IEMOCAP \cite{busso2008iemocap} (acted emotional dialogue) and ATCO2 \cite{zuluaga2023atco2} (noisy, jargon-dense air-traffic-control radio).
Each is subsampled to one hour in order to be manageable. All systems receive the same reference clips and use public checkpoints without fine-tuning: Qwen3-TTS (\texttt{Qwen3-TTS-12Hz-1.7B-Base}, 1.7\,B parameters), Chatterbox (500\,M), XTTS-v2 (${\sim}518$\,M) and F5-TTS v1 Base (336\,M). Qwen3-TTS runs in bfloat16 with sampling (temperature $0.9$, top-$k$ $50$, top-$p$ $1.0$, repetition penalty $1.05$); the baselines keep their libraries defaults hyper-parameters.


Qwen3-TTS offers two cloning interfaces: ICL, the default, receives the
reference audio with its transcript and continues in that voice, while
X-Vector conditions only on a speaker embedding \cite{snyder2018xvector}
and needs no transcript. We also vary the reference--target pairing. In
the aligned condition the model receives the audio and text of utterance
$i$ and renders $i$; in the shifted condition, which matches deployment,
the voice comes from utterance $i{+}1$ and the text from $i$, so prosody
cannot be copied. All results below use the shifted condition unless
stated otherwise.

\subsection{Perceptual test and LLM judge}
\label{sec:fooling-protocol}

We started by running a two-alternative forced-choice (2AFC) test
\cite{srinivasavaradhan2025hfr} on a web platform. Each round presents the real recording and its synthetic counterpart,
same text, as clips A and B in random order, listeners replay both
freely and pick the one judged real. The fooling rate is
the fraction of rounds in which the synthetic clip is chosen. Listeners are three entry-level annotators working at the same place,
so that familiarity with synthesis artefacts does not bias their
judgments.
For each system, 100 sampled pairs of AMI are split over
non-overlapping lists and used to compute the human fooling rate (Table~\ref{tab:human-fooling}). A second session, in which each listener judges 135 new pairs of AMI (new full 1~h subset) spread over
the whole combinations of 5 models and 2 conditions (shifted and aligned), and supply the data points used as the
development set in Section~\ref{sec:candidates}, for measure selection and correlation study. The full human annotation process took an estimated total of 12~hours to the annotators.

In order to extends these 2AFC study on all the others datasets, and at a more reproductible and reasonable cost, we query a multimodal LLM judge, capable of handle multiple audios in the same input, in our case, Gemini 3.5 Flash \cite{gemini35flash}. It answers exactly the same 100 pairs per system on AMI, and proceed to the same protocol on the 100 pairs from the other three corpora. It is queried at temperature $0$ with the answer constrained to the enum $\{\mathrm{A},\mathrm{B}\}$, and told only that one clip is real and one synthetic and to reply with the letter of the synthetic one. Because such judgments are sensitive to presentation order, every pair is judged twice, once in each ordering. The reported rate always uses the first ordering, with the real clip as A, and every pair stays in the denominator. The second ordering is a robustness diagnostic, and pairs whose verdict flips are counted as order-dependent and reported in Section~\ref{sec:rq5}.

Afterwards, listeners described in free form the cues behind their decisions, and their answers fall into four categories. All three mentioned background and channel sounds, such as background speakers or the radio beeps bracketing ATC transmissions, which occur only in real recordings; two noted breathing and microphone noise, absent from clones; all three reported differences in speaking rate and phrasing, since real speakers vary their rate within an utterance and follow punctuation whereas clones usually keep a steady rate; and two found clones flatter in emotional intensity on IEMOCAP.

\subsection{Objective measures}
\label{sec:candidates}

We compute 18 automatic measures comparing each synthetic clip with its real
counterpart, in four dimensions. \texttt{Speaker}: a ResNet34 embedding
cosine \cite{wang2023wespeaker}, vocal-burst rate and class match from a
VoiceCLAP-based detector of 82 non-verbal vocalisations
\cite{vocalburst}, an emotion2vec cosine \cite{ma2024emotion2vec} and
SpeechBERTScore \cite{saeki2024speechbertscore} on WavLM-Large layer-14
features. \texttt{Prosody}: relative duration error, speech-ratio
difference from an energy-based VAD, gain-variation match ($\sigma/\mu$ of
frame RMS) and mean frame-RMS difference. \texttt{$F_0$}: per-utterance $F_0$ using vocoder estimator from \cite{saeki2024speechbertscore}, then mean and standard-deviation differences are applied and a DTW-aligned log-$F_0$ RMSE. \texttt{Intelligibility}: WER difference between
Whisper Large V3 \cite{radford2023whisper} transcripts of the synthetic
and real clips, DNSMOS P.835 \cite{reddy2022dnsmos} BAK and OVRL
differences, and STOI, PESQ and reference-free SI-SDR by
TorchAudio-SQUIM \cite{kumar2023squim}.

We combine these automatic measures into four dimension scores and metric filtering is based
on the correlation between the metrics and human annotations on the development set kept separate. We retain from that 13 measures, whose for which agreement with human fooling validate a Spearman correlation $\rho\ge0.45$ on $n=4$. Other measures are dropped as no strong signal is carried into the human perception. The retained set is then applied to data it was not selected on, including the 3 other systems and 3 other corpora with both conditions (shifted and aligned).

Each retained measure is first mapped to a per-utterance agreement $A\in[0,1]$, $1$ meaning that the synthetic clip matches its real counterpart within a tolerance cap set per measure, so that dimensionless and unbounded quantities become comparable. Let $\mathcal{C}=\{\mathrm{Spk},\mathrm{Pro},F_0,\mathrm{Int}\}$ and let $A_{c,i}$ be the agreement of the $i$-th of the $n_c$ measures retained in dimension $c$. Each measure is mapped to a reliability, averaged within its dimension, and the same map is applied again to that dimension mean, giving the four dimension scores $S_c$ and their unweighted mean $S$ on a $0$--$100$ scale:

\begin{equation}
\begin{aligned}
r_{c,i} &= 1-(1-A_{c,i})^{2}, &
\bar r_c &= \tfrac{1}{n_c}\textstyle\sum_{i} r_{c,i},\\[2pt]
S_c &= 1-(1-\bar r_c)^{2}, &
S &= \tfrac{100}{|\mathcal{C}|}\textstyle\sum_{c\in\mathcal{C}} S_c .
\end{aligned}
\label{eq:score}
\end{equation}

The double mapping keeps the scale from saturating as systems approach human quality, where raw agreements crowd towards $1$. Weighting the four dimensions equally is a deliberate choice rather than a calibrated one, and Section~\ref{sec:rq5} tests it against human judgments rather than assuming it.




\subsection{Hallucination estimation}
\label{sec:hall-estimation}

Hallucinated content, meaning repeated, skipped or invented words, is
common in language-model TTS \cite{song2025ellav}. We flag it from
timing alone, which keeps the test applicable at dataset scale and at
controlled computational cost. Writing $d$ for the duration of the
synthetic clip, $n_w$ for the number of words in its target text and
$d_{\mathrm{ref}}$ for the duration of the real counterpart, a clip is
flagged when
\begin{equation}
h = \mathbf{1}\!\left[\; \frac{d}{n_w} > \tau_w
      \;\;\lor\;\; \frac{d}{d_{\mathrm{ref}}} > 1+\tau_d \;\right],
\qquad \tau_w = 1\,\mathrm{s},\; \tau_d = 0.5 .
\label{eq:halluc}
\end{equation}
The first term catches clips that stall or loop on few words, the second
those that run past their reference, the two duration signatures of
repeated or invented content. Because the rule reads duration rather
than lexical content, two authors audited $100$ flagged and $100$
unflagged utterances across systems and corpora. It misses few
hallucinations, $3$ of $57$ (recall $94.7\%$, $F_1$ $68.8\%$), and
unflagged clips are rarely hallucinated ($3$ of $100$).
Flagged rates therefore overestimate the true hallucination rate, but the same rule applies to every system and corpus, so comparisons between them remain valid.

\subsection{Spoofing countermeasure}
\label{sec:aasist}

A clone that convinces a listener may still be trivially detectable by
machine, so we submit the same audios to AASIST \cite{jung2022aasist},
an end-to-end raw-waveform anti-spoofing model. We take the released
checkpoint\footnote{\url{https://github.com/clovaai/aasist}} ($0.30$\,M
parameters) trained on the logical-access partition of ASVspoof~2019
\cite{todisco2019asvspoof} and apply it exactly as distributed, with no
retraining or threshold recalibration, so that it stands for the
off-the-shelf countermeasure a practitioner would reach for rather than
the best achievable detector. Audio is converted to $16$\,kHz mono and truncated or zero-padded to
$64{,}600$ samples, the original protocol's fixed length. A
clip is labelled spoof whenever the spoof logit exceeds the bona-fide
logit. Both clips of a pair are preprocessed identically. Each cell of
Table~\ref{tab:deepfake-detection} covers 100 shifted pairs. We report the detection rate (Det.) which consist of the percentage of clones identified properly as spoof, and the false-alarm rate (FA) which is the percentage of their genuine counterparts miss-identified as spoof.

\section{Results}
\label{sec:results}

\subsection{Perceived realism and alternatives}
\label{sec:rq1}

Qwen3-TTS is the hardest of the five systems to tell from real speech, but it remains detectable. On AMI, listeners pick its clones as real in $35.6\%$ (X-Vector) and $27.0\%$ (ICL) of rounds, against at most $9.0\%$ for the baselines (Table~\ref{tab:human-fooling}).
This is consistent with the difficulty listeners have with other modern voice cloning systems \cite{mai2023warning,barrington2025voiceclones}.
Their cues (Section~\ref{sec:fooling-protocol}) show what gives clones away: they reproduce the voice but not the sounds around it, nor the rate variation and emotional intensity of natural speech.

\begin{table}[H]
\centering
\caption{2AFC fooling rate (\%) on 100 randomly sampled utterances per
system and corpus, shifted condition.}
\label{tab:human-fooling}
\vspace{1mm}
\scriptsize
\setlength{\tabcolsep}{3.5pt}
\resizebox{\columnwidth}{!}{%
\begin{tabular}{lccccc}
\toprule
& \multicolumn{2}{c}{\textbf{AMI}} & \textbf{IEMOCAP} & \textbf{ATCO2} & \textbf{LS} \\
\cmidrule(lr){2-3}\cmidrule(lr){4-4}\cmidrule(lr){5-5}\cmidrule(lr){6-6}
\textbf{System} & \textbf{Human} & \textbf{Gemini} & \textbf{Gemini} & \textbf{Gemini} & \textbf{Gemini} \\
\midrule
Qwen3 (ICL)   & $27.0$ & $12.0$ & $15.0$ & $6.0$ & $14.0$ \\
Qwen3 (X-Vec) & $35.6$ & $11.0$ & $7.0$ & $3.0$  & $10.0$ \\
\midrule
Chatterbox    & $9.0$ & $11.0$ & $10.0$ & $1.0$  & $9.0$ \\
XTTS-v2       & $6.0$ & $3.0$ & $3.0$ & $4.0$   & $26.0$\\
F5-TTS        & $1.0$ & $5.0$ & $5.0$ & $5.0$  & $7.0$ \\
\bottomrule
\end{tabular}
}
\end{table}

On LibriSpeech, XTTS-v2 fools the judge most often ($26\%$) and Chatterbox reaches the highest objective score of any system ($S=51.9$, Table~\ref{tab:score-systems}). Outside read speech that advantage disappears. XTTS-v2 collapses on ATCO2 and IEMOCAP ($S=11.6$ and $12.9$) and hallucinates on $18.5\%$ of its samples, almost all on those two short-utterance corpora, and F5-TTS is the weakest system overall ($S=21.1$). Chatterbox is the most robust alternative, close to Qwen3-TTS on average ($31.9$ vs.\ $33.4$--$34.3$) and best on prosody, yet it fools our listeners only $9\%$ of the time.


\subsection{Machine detection}
\label{sec:rq1b}

A standard detector does not fill the gap left by listeners, and on Qwen3-TTS it barely works at all (Table~\ref{tab:deepfake-detection}). Since the false-alarm rate varies by corpus, a system is detectable only where its detection rate clearly exceeds it. By that criterion Qwen3-TTS is detectable in one of the eight cells measured, ICL on LibriSpeech ($39.0\%$ against $19.0\%$); in the other seven the detector flags the real recordings at least as often as the clones. X-Vector on AMI is the extreme case, caught on one clip in a hundred while a sixth of the genuine meeting recordings are rejected.

Detectability is a property of the generator rather than of perceived quality, and the two orderings do not match: XTTS-v2 is the only system the detector separates convincingly ($86$--$91\%$ in all four corpora) yet is among the least convincing for our listeners ($6\%$ on AMI), while F5-TTS, the least convincing of all ($1\%$), is caught on $36.0\%$ of AMI clips against $14.0\%$ false alarms. A clone a detector recognises is therefore not a clone a listener recognises.

On ATC radio the failure is complete and system-independent, the
detector flags $99\%$ of the \emph{genuine} ATCO2 recordings as spoof,
more often than the clones of any system ($75$--$95\%$), including the
XTTS-v2 clones it identifies almost perfectly everywhere else. Its
decision is driven by the narrow-band, noisy, clipped VHF channel,
nothing like the studio-quality bona-fide speech of ASVspoof~2019
\cite{todisco2019asvspoof}, and not by any property of the clone; this
is the generalisation gap reported for deployed anti-spoofing
\cite{gengap2025} in an extreme form, and no operating point recovers a
discrimination the scores do not contain. Off-the-shelf anti-spoofing thus offers no protection for ATC, a
safety-critical domain in which a spoofed transmission is exactly the
attack that matters.

\begin{table}[H]
  \centering
  \caption{AASIST detection and false-alarm rates (\%) per corpus and
  system, shifted condition. Bold: detection
  clearly above the paired false-alarm rate.}
  \label{tab:deepfake-detection}
  \vspace{1mm}
  \resizebox{\columnwidth}{!}{%
  \setlength{\tabcolsep}{3pt}
  \renewcommand{\arraystretch}{0.95}
  \begin{tabular}{@{}ll ccccc@{}}
    \toprule
    \multirow{2}{*}{\textbf{Corpus}} & & Qwen3 & Qwen3
      & \multirow{2}{*}{Chatterbox}
      & \multirow{2}{*}{XTTS-v2}
      & \multirow{2}{*}{F5-TTS} \\
    & & (ICL) & (X-Vec) &  & & \\
    \midrule
    \multirow{2}{*}{AMI}
      & Det. & $13.0$ & $1.0$ & $\mathbf{24.0}$ & $\mathbf{89.0}$ & $\mathbf{36.0}$ \\
      & FA   & $17.0$ & $17.0$ & $19.0$ & $17.0$ & $14.0$ \\
    \midrule
    \multirow{2}{*}{IEMOCAP}
      & Det. & $10.0$ & $8.0$ & $12.0$ & $\mathbf{91.0}$ & $11.0$ \\
      & FA   & $14.0$ & $14.0$ & $14.0$ & $14.0$ & $14.0$ \\
    \midrule
    \multirow{2}{*}{ATCO2}
      & Det. & $92.0$ & $87.0$ & $95.0$ & $91.0$ & $75.0$ \\
      & FA   & $99.0$ & $99.0$ & $99.0$ & $99.0$ & $99.0$ \\
    \midrule
    \multirow{2}{*}{LS}
      & Det. & $\mathbf{39.0}$ & $6.0$ & $16.0$ & $\mathbf{86.0}$ & $\mathbf{20.0}$ \\
      & FA   & $19.0$ & $19.0$ & $19.0$ & $19.0$ & $19.0$ \\
    \bottomrule
  \end{tabular}
  }
\end{table}

\subsection{In-context learning or X-Vector?}
\label{sec:rq2}

Humans on AMI are fooled more by X-Vector
($35.6\%$ vs.\ $27.0\%$), a gap too small for our panel to settle,
whereas the LLM judge prefers ICL on all four corpora
(Table~\ref{tab:human-fooling}). Their objective profiles explain why. ICL clones timbre better from
clean references ($S_{\mathrm{Spk}}$ $80.2$ vs.\ $61.5$ on LibriSpeech,
$52.2$ vs.\ $43.6$ on AMI), while X-Vector produces closer prosody on
every corpus ($S_{\mathrm{Pro}}$ higher by $2.6$ to $11.3$ points) and
matches or beats ICL on speaker similarity for noisy ATCO2 and short
IEMOCAP references. Overall scores are nearly tied ($33.4$ vs.\ $34.3$).
X-Vector is also safer, since ICL hallucinates $1.6$ times as often
($4.1\%$ vs.\ $2.5\%$) because it must reconcile the reference
transcript with the audio (Section~\ref{sec:rq3}).

\subsection{Domains and failure modes}
\label{sec:rq3}

\begin{figure}[H]
  \centering
  \includegraphics[width=0.99\columnwidth]{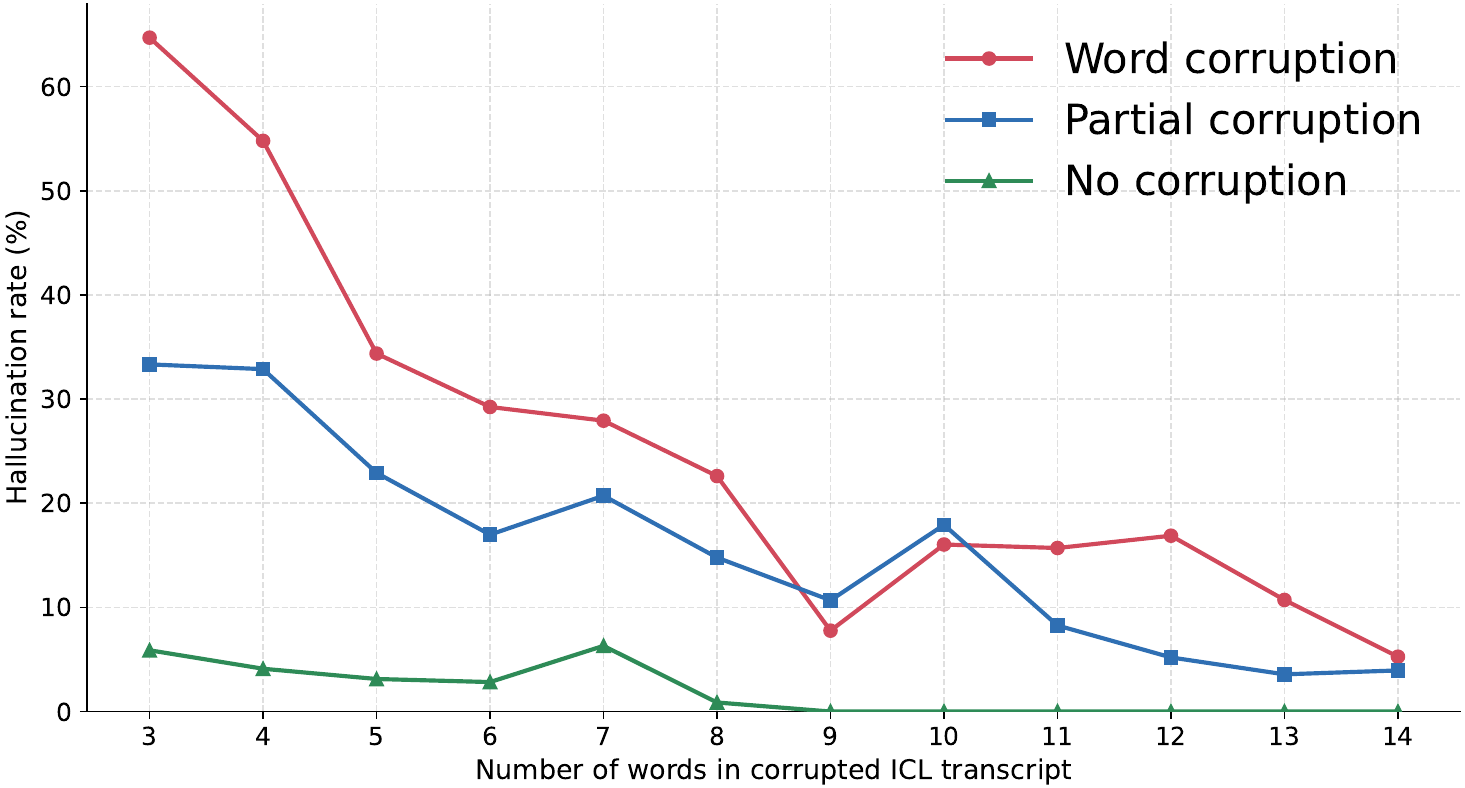}
  \caption{Hallucination rate of Qwen3-TTS ICL against reference transcript length (shifted condition).}
  \label{fig:hallucinations-values_corruption}
\end{figure}

In a two-way decomposition of the scores of Table~\ref{tab:score-systems}, corpus
explains $63\%$ of the variance and system only $15\%$. Showing that domains matter more than the choice of system. LibriSpeech is
easiest (mean $S=43.4$) and ATCO2 hardest ($18.5$), where
intelligibility collapses for every system ($S_{\mathrm{Int}}$ $14.7$
vs.\ $55.2$) and the judge is fooled in only $3.8\%$ of rounds.
Emotional speech like IEMOCAP has the lowest $F_0$ and prosody scores ($10.5$,
$19.2$), since expressive pitch and timing are hard to transfer from a
short reference. Listeners single out the same two domains.


We corrupt one word in the middle of it while keeping the audio the same. Partial corruption (PC) slightly alters that word, word corruption (WC) replaces it with a static distractor. The clearest failure mode of Qwen3-TTS is an unreliable reference transcript under ICL.  On the short-utterance corpora the hallucination rate rises by an order of magnitude (IEMOCAP $0.8\%\!\to\!21.2\%$ under WC, ATCO2 $1.6\%\!\to\!15.2\%$, means of $11.2$ and $12.2$ words per utterance), while the long AMI and LS references ($84.2$ and $33.8$ words) stay below $1.5\%$; pooled over corpora the rate goes from $1.1\%$ to $9.0\%$ under PC and $14.4\%$ under WC.
Transcript length, not corpus, is the driver: the WC rate falls from
$65\%$ on three-word transcripts to below $20\%$ once the transcript
exceeds eight words, and PC follows the same trend from $33\%$
(Fig.~\ref{fig:hallucinations-values_corruption}). With few words to
align, a single wrong word dominates the conditioning, while diluted in longer
context.


Transcribing the corrupted syntheses with Whisper Large V3 reveals a mechanism more specific than a generic loss of alignment. The distractor is audibly reproduced in $7.0\%$ of utterances, again concentrated on the short-utterance corpora ($12.2\%$ on IEMOCAP, $6.1\%$ on ATCO2, against $1.8\%$ and $0.7\%$ elsewhere), and where it leaks it almost never appears where it was injected (mid-sentence) but rather, it surfaces in the first third of the words of the synthesis in $96.6\%$ of cases and as the literal first word in $57\%$, never in the final third. The model thus emits the unexpected word, due to a attention leak, while still anchoring on the reference, consistent with a continuation-style hallucinations before committing to the target.

\begin{table}[H]
\centering
\caption{Overall score $S$ and its four dimension scores per system and corpus, shifted condition.}
\label{tab:score-systems}
\vspace{1mm}
\setlength{\tabcolsep}{4pt}
\renewcommand{\arraystretch}{0.88}
\resizebox{\columnwidth}{!}{%
\begin{tabular}{@{}ll rrrr r@{}}
\toprule
\textbf{System} & \textbf{Corpus}
 & $S_{\mathrm{Spk}}$ & $S_{\mathrm{Pro}}$ & $S_{F_0}$ & $S_{\mathrm{Int}}$ & $S$ \\
\midrule
\multirow{4}{*}{Qwen3 (ICL)}
 & AMI         & $52.2$ & $15.7$ & $25.2$ & $50.8$ & $36.0$ \\
 & ATCO2       & $31.6$ & $21.5$ & $10.6$ & $15.4$ & $19.8$ \\
 & LibriSpeech & $80.2$ & $33.2$ & $21.7$ & $64.0$ & $49.8$ \\
 & IEMOCAP     & $32.9$ & $25.9$ & $15.0$ & $38.2$ & $28.0$ \\
\midrule
\multirow{4}{*}{Qwen3 (X-Vector)}
 & AMI         & $43.6$ & $23.7$ & $26.6$ & $53.3$ & $36.8$ \\
 & ATCO2       & $34.4$ & $24.1$ & $15.0$ & $15.9$ & $22.4$ \\
 & LibriSpeech & $61.5$ & $42.4$ & $20.4$ & $67.1$ & $47.9$ \\
 & IEMOCAP     & $34.5$ & $37.2$ & $11.7$ & $37.7$ & $30.3$ \\
\midrule
\multirow{4}{*}{Chatterbox}
 & AMI         & $24.4$ & $26.2$ & $21.1$ & $40.9$ & $28.1$ \\
 & ATCO2       & $16.1$ & $32.3$ & $14.6$ & $16.9$ & $20.0$ \\
 & LibriSpeech & $81.6$ & $47.7$ & $20.4$ & $57.8$ & $51.9$ \\
 & IEMOCAP     & $40.3$ & $29.1$ & $8.8$  & $31.6$ & $27.4$ \\
\midrule
\multirow{4}{*}{XTTS-v2}
 & AMI         & $59.4$ & $44.7$ & $13.9$ & $45.3$ & $40.9$ \\
 & ATCO2       & $19.8$ & $2.1$  & $9.8$  & $14.7$ & $11.6$ \\
 & LibriSpeech & $54.0$ & $54.5$ & $19.2$ & $61.9$ & $47.4$ \\
 & IEMOCAP     & $15.4$ & $2.7$  & $5.1$  & $28.4$ & $12.9$ \\
\midrule
\multirow{4}{*}{F5-TTS}
 & AMI         & $64.7$ & $9.3$  & $22.4$ & $19.5$ & $29.0$ \\
 & ATCO2       & $29.6$ & $19.1$ & $14.8$ & $10.4$ & $18.5$ \\
 & LibriSpeech & $20.6$ & $19.1$ & $15.2$ & $25.3$ & $20.0$ \\
 & IEMOCAP     & $34.0$ & $1.2$  & $11.7$ & $21.0$ & $17.0$ \\
\midrule
\midrule
\multicolumn{2}{@{}l}{Average, Qwen3 (ICL)}      & $49.2$ & $24.1$ & $18.1$ & $42.1$ & $33.4$ \\
\multicolumn{2}{@{}l}{Average, Qwen3 (X-Vector)} & $43.5$ & $31.9$ & $18.4$ & $43.5$ & $34.3$ \\
\multicolumn{2}{@{}l}{Average, Chatterbox}       & $40.6$ & $33.8$ & $16.2$ & $36.8$ & $31.9$ \\
\multicolumn{2}{@{}l}{Average, XTTS-v2}          & $37.2$ & $26.0$ & $12.0$ & $37.6$ & $28.2$ \\
\multicolumn{2}{@{}l}{Average, F5-TTS}           & $37.2$ & $12.2$ & $16.0$ & $19.0$ & $21.1$ \\
\bottomrule
\end{tabular}
}
\end{table}

\subsection{Acoustic strengths and weaknesses}
\label{sec:rq4}

Table~\ref{tab:score-systems} shows where Qwen3-TTS stands out. One of its two modes reaches the highest $S_{\mathrm{Int}}$ on three of the four corpora and the highest $S_{F_0}$ on all four, and it reproduces each gender's pitch register, the per-utterance mean $F_0$ of ICL clones tracking the real distribution with medians only a few Hz low. Its average speaker similarity under ICL is the best of the five, though not in every domain, since F5-TTS leads on AMI and Chatterbox on LibriSpeech and IEMOCAP. Prosody is its main weakness, with Chatterbox and XTTS-v2 closer to the real timing on three corpora. In absolute terms $F_0$ is the weakest dimension for every system but F5-TTS ($12.0$ to $18.4$), since pitch registers match at the population level while each clone drifts from the specific speaker it should reproduce.

\subsection{Can automatic measures replace listeners?}
\label{sec:rq5}

Over the development set, background and channel sounds are well captured by DNSMOS background differences ($\rho=0.77$), breathing only partly by the vocal-burst rate ($0.52$) and not by which bursts occur ($0.32$), and speaking rate partly by duration error and VAD ratio ($0.71$, $0.54$), which capture global timing but not rate changes within an utterance or where pauses fall. Emotional intensity is missed entirely: emotion2vec ($0.11$), gain variation ($0.17$) and mean energy ($0.37$) barely track fooling, and SpeechBERTScore carries no signal ($0.01$). The strongest correlate, SQUIM SI-SDR ($0.81$), is reference-free, so it partly tracks how clean a corpus is rather than how a clone differs from its reference.

Averaged over the four corpora, $S$ ranks the systems exactly as our listeners do on AMI: Qwen3-TTS first ($34.3$, $33.4$), then Chatterbox ($31.9$), XTTS-v2 ($28.2$) and F5-TTS ($21.1$). The agreement does not survive a matched comparison. On AMI alone $S$ puts XTTS-v2 first ($40.9$), yet it fools humans in only $6\%$ of rounds, and over the five systems $S$ correlates weakly with fooling ($\rho=0.20$) while speaker similarity correlates negatively ($-0.70$), F5-TTS having the most similar voice ($S_{\mathrm{Spk}}=64.7$) and the least convincing clone; only intelligibility and quality follows listeners ($0.90$). Averaging reproduces the human ordering by mixing in domain difficulty, not by tracking what listeners hear.

The LLM judge fails differently, on AMI it agrees with the human on at least $78\%$ of Chatterbox, XTTS-v2 and F5-TTS clones, but on under $73\%$ of Qwen3-TTS clones and on at most $16.7\%$ of the utterances that fooled the human. It is also unstable, even with deterministic decoding, the two presentation orders (Section~\ref{sec:fooling-protocol}) disagree on up to $43\%$ of pairs (XTTS-v2, LibriSpeech). 
\section{Discussion and Conclusion}
\label{sec:conclusion}

With Qwen3-TTS, the available reference should decide the interface, since neither dominates. ICL clones timbre more closely and X-Vector timing, so ICL suits clean references whose transcript is verified and over ten words, X-Vector short, noisy or untranscribed ones and wherever timing matters most. The domain decides the rest, explaining four times more variance than the system, so a short in-domain validation is worth more than switching models. Between systems, Qwen3-TTS leads on intelligibility and pitch register and is hardest to tell from real speech, yet lags on prosody and per-speaker pitch. Chatterbox is a robust second, and XTTS-v2 should be restricted to read speech. No proxy settles that choice, because neither our objective score nor an LLM judge reproduces the listener ranking. No measure we tested tracks perceived realism closely enough to replace a listener, so automatic evaluation of cloned speech is an open problem and a small forced-choice test in the target domain the only arbiter. Safeguards fare no better, as an off-the-shelf countermeasure flags Qwen3-TTS clones no more often than genuine recordings outside read speech and is unusable on ATC radio at any threshold, making disclosure \cite{euaiact2024} more dependable than post-hoc detection. 
Future work should target ATC, where a cloned controller voice on a live frequency is a safety risk that no detector we tested can flag, making a channel-matched countermeasure the priority.

\newpage

\section{Acknowledgments}
Idiap employees were funded by the European Union's Horizon 2020 project
ELOQUENCE (grant 101070558). Séverin Baroudi and Ricard Marxer were
supported by the Agence de l'Innovation de Défense (grant 2022~65~0079),
the French National Research Agency (ANR-20-CE23-0012-01, MIM), and
computational resources from GENCI-IDRIS (grant AD011014044R2).

\section{Compliance with Ethical Standards}
This study uses speech corpora that are openly available or distributed
under institutional licence (LibriSpeech, AMI, IEMOCAP, ATCO2). The listening test was answered by
adult volunteers who gave informed consent, and only their anonymous
responses were recorded. No personal data were collected.

\bibliographystyle{IEEEbib}
\bibliography{refs-min}

\begin{thebibliography}{10}

\bibitem{wang2023valle}
Sanyuan Chen, Chengyi Wang, Yu~Wu, et~al.,
\newblock ``Neural codec language models are zero-shot text to speech synthesizers,''
\newblock {\em IEEE Trans. Audio, Speech, Lang. Process.}, vol. 33, pp. 705--718, 2025.

\bibitem{casanova2024xtts}
Edresson Casanova, Kelly Davis, Eren G{\"o}lge, et~al.,
\newblock ``{XTTS}: A massively multilingual zero-shot text-to-speech model,''
\newblock in {\em Proc. Interspeech}, 2024, pp. 4978--4982.

\bibitem{chen2024f5}
Yushen Chen, Zhikang Niu, Ziyang Ma, et~al.,
\newblock ``{F5-TTS}: A fairytaler that fakes fluent and faithful speech with flow matching,''
\newblock in {\em Proc. ACL}, 2025, pp. 6255--6271.

\bibitem{qwen3tts}
Hangrui Hu, Xinfa Zhu, Ting He, et~al.,
\newblock ``{Qwen3-TTS} technical report,'' 2026,
\newblock arXiv:2601.15621.

\bibitem{chen2024valle2}
Sanyuan Chen, Shujie Liu, Long Zhou, et~al.,
\newblock ``{VALL-E 2}: Neural codec language models are human parity zero-shot text to speech synthesizers,'' 2024,
\newblock arXiv:2406.05370.

\bibitem{ju2024naturalspeech3}
Zeqian Ju, Yuancheng Wang, Kai Shen, et~al.,
\newblock ``{NaturalSpeech 3}: Zero-shot speech synthesis with factorized codec and diffusion models,''
\newblock in {\em Proc. ICML}, 2024.

\bibitem{radford2023whisper}
Alec Radford, Jong~Wook Kim, Tao Xu, et~al.,
\newblock ``Robust speech recognition via large-scale weak supervision,''
\newblock in {\em Proc. ICML}, 2023.

\bibitem{snyder2018xvector}
David Snyder, Daniel Garcia-Romero, Gregory Sell, et~al.,
\newblock ``X-vectors: Robust {DNN} embeddings for speaker recognition,''
\newblock in {\em Proc. IEEE ICASSP}, 2018, pp. 5329--5333.

\bibitem{wang2023wespeaker}
Hongji Wang, Chengdong Liang, Shuai Wang, et~al.,
\newblock ``{WeSpeaker}: A research and production oriented speaker embedding learning toolkit,''
\newblock in {\em Proc. IEEE ICASSP}, 2023, pp. 1--5.

\bibitem{reddy2022dnsmos}
Chandan K.~A. Reddy, Vishak Gopal, and Ross Cutler,
\newblock ``{DNSMOS P.835}: A non-intrusive perceptual objective speech quality metric to evaluate noise suppressors,''
\newblock in {\em Proc. IEEE ICASSP}, 2022, pp. 886--890.

\bibitem{lemaguer2024mos}
S{\'e}bastien~Le Maguer, Simon King, and Naomi Harte,
\newblock ``The limits of the mean opinion score for speech synthesis evaluation,''
\newblock {\em Comput. Speech Lang.}, vol. 84, pp. 101577, 2024.

\bibitem{minixhofer2025ttsds2}
Christoph Minixhofer, Ond{\v r}ej Klejch, and Peter Bell,
\newblock ``{TTSDS2}: Robust objective evaluation for human-quality synthetic speech,''
\newblock in {\em Proc. ISCA SSW}, 2025, pp. 68--75.

\bibitem{srinivasavaradhan2025hfr}
Praveen {Srinivasa Varadhan}, Sherry Thomas, Sai {Teja M S}, et~al.,
\newblock ``The state of {TTS}: A case study with human fooling rates,''
\newblock in {\em Proc. Interspeech}, 2025, pp. 2285--2289.

\bibitem{chiang2025judges}
Cheng-Han Chiang, Xiaofei Wang, Chung-Ching Lin, et~al.,
\newblock ``Audio-aware large language models as judges for speaking styles,''
\newblock in {\em Proc. EMNLP Findings}, 2025, pp. 467--480.

\bibitem{zhang2026speechjudge}
Xueyao Zhang, Chaoren Wang, Huan Liao, et~al.,
\newblock ``{SpeechJudge}: Towards human-level judgment for speech naturalness,''
\newblock in {\em Proc. ICLR}, 2026.

\bibitem{yamagishi2021asvspoof}
Junichi Yamagishi, Xin Wang, Massimiliano Todisco, et~al.,
\newblock ``{ASVspoof 2021}: Accelerating progress in spoofed and deepfake speech detection,''
\newblock in {\em Proc. ASVspoof Challenge Workshop}, 2021, pp. 47--54.

\bibitem{jung2022aasist}
Jee weon Jung, Hee-Soo Heo, Hemlata Tak, et~al.,
\newblock ``{AASIST}: Audio anti-spoofing using integrated spectro-temporal graph attention networks,''
\newblock in {\em Proc. IEEE ICASSP}, 2022, pp. 6367--6371.

\bibitem{mai2023warning}
Kimberly~T. Mai, Sergi Bray, Toby Davies, et~al.,
\newblock ``Warning: Humans cannot reliably detect speech deepfakes,''
\newblock {\em PLOS ONE}, vol. 18, no. 8, pp. 1--20, 2023.

\bibitem{barrington2025voiceclones}
Sarah Barrington, Emily~A. Cooper, and Hany Farid,
\newblock ``People are poorly equipped to detect {AI}-powered voice clones,'' 2025,
\newblock arXiv:2410.03791.

\bibitem{lavan2025hyperreal}
Nadine Lavan, Mairi Irvine, Victor Rosi, et~al.,
\newblock ``Voice clones sound realistic but not (yet) hyperrealistic,''
\newblock {\em PLOS ONE}, vol. 20, no. 9, pp. 1--19, 2025.

\bibitem{gengap2025}
Victoria~Garc{\'i}a Mart{\'i}nez-Echevarr{\'i}a, Rafael Palacios, Gregorio L{\'o}pez, et~al.,
\newblock ``The generalization gap: Do audio deepfake detectors actually protect against modern vishing?,''
\newblock {\em Electronics}, vol. 15, no. 13, pp. 2846, 2026.

\bibitem{euaiact2024}
``Regulation ({EU}) 2024/1689 ({AI Act}), art.~50,'' Off. J. Eur. Union, 2024.

\bibitem{panayotov2015librispeech}
Vassil Panayotov, Guoguo Chen, Daniel Povey, et~al.,
\newblock ``{LibriSpeech}: An {ASR} corpus based on public domain audio books,''
\newblock in {\em Proc. IEEE ICASSP}, 2015, pp. 5206--5210.

\bibitem{carletta2005ami}
Jean Carletta, Simone Ashby, Sebastien Bourban, et~al.,
\newblock ``The {AMI} meeting corpus: A pre-announcement,''
\newblock in {\em Proc. MLMI}, 2005, pp. 28--39.

\bibitem{busso2008iemocap}
Carlos Busso, Murtaza Bulut, Chi-Chun Lee, et~al.,
\newblock ``{IEMOCAP}: Interactive emotional dyadic motion capture database,''
\newblock {\em Lang. Resour. Eval.}, vol. 42, no. 4, pp. 335--359, 2008.

\bibitem{zuluaga2023atco2}
Juan Zuluaga-Gomez, Karel Vesel{\'y}, Igor Sz{\"o}ke, et~al.,
\newblock ``{ATCO2} corpus: A large-scale dataset for research on automatic speech recognition and natural language understanding of air traffic control communications,'' 2022,
\newblock arXiv:2211.04054.

\bibitem{gemini35flash}
{Google DeepMind},
\newblock ``{Gemini 3.5 Flash} model card,'' \url{deepmind.google/models/model-cards/gemini-3-5-flash/}, 2026.

\bibitem{vocalburst}
{LAION},
\newblock ``Vocal burst detector v2,'' Hugging Face model card, \url{hf.co/laion/vocal-burst-detector-v2}, 2026.

\bibitem{ma2024emotion2vec}
Ziyang Ma, Zhisheng Zheng, Jiaxin Ye, et~al.,
\newblock ``emotion2vec: Self-supervised pre-training for speech emotion representation,''
\newblock in {\em Proc. ACL Findings}, 2024, pp. 15747--15760.

\bibitem{saeki2024speechbertscore}
Takaaki Saeki, Soumi Maiti, Shinnosuke Takamichi, et~al.,
\newblock ``{SpeechBERTScore}: Reference-aware automatic evaluation of speech generation leveraging {NLP} evaluation metrics,''
\newblock in {\em Proc. Interspeech}, 2024, pp. 4943--4947.

\bibitem{kumar2023squim}
Anurag Kumar, Ke~Tan, Zhaoheng Ni, et~al.,
\newblock ``{Torchaudio-Squim}: Reference-less speech quality and intelligibility measures in {Torchaudio},''
\newblock in {\em Proc. IEEE ICASSP}, 2023, pp. 1--5.

\bibitem{song2025ellav}
Yakun Song, Zhuo Chen, Xiaofei Wang, et~al.,
\newblock ``{ELLA-V}: Stable neural codec language modeling with alignment-guided sequence reordering,''
\newblock in {\em Proc. AAAI}, 2025.

\bibitem{todisco2019asvspoof}
Massimiliano Todisco, Xin Wang, Ville Vestman, et~al.,
\newblock ``{ASVspoof 2019}: Future horizons in spoofed and fake audio detection,''
\newblock in {\em Proc. Interspeech}, 2019, pp. 1008--1012.

\end{thebibliography}

\end{document}